\pdfoutput=1

\documentclass{article} 

\usepackage[utf8]{inputenc} 
\usepackage{hyperref}
\hypersetup{
	linktocpage,
	breaklinks=true,   
	colorlinks=true,   
	pdfusetitle=true,  
}

\usepackage[
backend=bibtex,
sorting=ynt
]{biblatex}
\bibliography{doublecheck}

\usepackage{geometry} 
\usepackage{graphicx} 
\usepackage{setspace}

\usepackage{booktabs} 
\usepackage{array} 
\usepackage{paralist} 
\usepackage{verbatim} 
\usepackage{subfig} 

\usepackage{fancyhdr} 
\usepackage{sectsty}

\usepackage[nottoc,notlof,notlot]{tocbibind} 
\usepackage[titles,subfigure]{tocloft} 
\usepackage{amsmath,amssymb,amsthm}
\usepackage{color}

\title{An Efficient Modular Exponentiation Proof Scheme}
\author{Darren Li, Yves Gallot}
\date{} 

\begin{document}
	\maketitle
	{
		
		\begin{abstract}
			We present an efficient proof scheme for any instance of left-to-right modular exponentiation, used in many computational tests for primality. Specifically, we show that for any $(a,n,r,m)$ the correctness of a computation $a^n\equiv r\pmod m$ can be proven and verified with an overhead negligible compared to the computational cost of the exponentiation. Our work generalizes the Gerbicz-Pietrzak proof scheme used when $n$ is a power of $2$, and has been successfully implemented at PrimeGrid, doubling the efficiency of distributed searches for primes.
		\end{abstract}
	}
	{
		\tableofcontents
	}
	
	\section{Introduction}
	
	In distributed computing and particularly volunteer computing, the task of efficiently discerning correct results from incorrect results, whether malicious or due to sheer chance, is notoriously difficult; it is necessary to suspect all results as possibly incorrect, and these suspicions can only be settled with peer verification. In most cases, such verification would take the form of an entire recomputation, doubling the amount of necessary computational power to complete the same amount of work, all for a result that is most likely correct.
	
	The majority of ongoing organized searches for primes of record-breaking sizes, such as PrimeGrid \cite{primegrid}, are distributed. A significant portion of computing power is thus wasted due to the inefficient verification process, where a single computation is done twice to ensure the correctness of every result.
	
	For a candidate prime $m$, the bottleneck of the Fermat probable prime test lies in evaluating $a^{m-1} \pmod m$ or a nearby power, requiring $\tilde{O}(\log^2 m)$ exact computations, where a single precision error will render the rest of the calculation incorrect. To prevent these errors that could potentially categorize a prime as composite, verification of all results are necessary.
	
	With the development of Krzysztof Pietrzak's verifiable delay function (VDF) \cite{pietrzak:LIPIcs:2018:10153} and in turn the Gerbicz-Pietrzak proof scheme \cite{atnashev}, when the exponent $n$ is a power of $2$ it became possible to create ``certificates'' -- non-interactive proofs of the result of modular exponentiation -- with minimal overhead, that could also be verified much faster than recomputing the test. However, it relies on the fact that for such $n$, the computation consists entirely repeated squaring, which is not true in the general case.
	
	We present a proof scheme for the modular exponentiation process suitable for \emph{any} input, allowing any instance of modular exponentiation to be verified much quicker and halving the total work necessary to conduct a primality test on a distributed system. This proof scheme has doubled the efficiency of the current search for Generalized Fermat primes, a form introduced for its unprecedentedly fast algorithms and GPU implementations \cite{genefer}.
	
	\subsection{Previous work}
	
	The Gerbicz-Pietrzak proof scheme, initially derived as an error check by Robert Gerbicz and modified to a proof scheme by Pavel Atnashev using Pietrzak's VDF \cite{atnashev}, can only be used when the desired result can be derived from $a^{n}$ where $n$ is a power of $2$. This is only the case for very specific instances of modular exponentiation, such as the primality test for Proth primes.
	
	The Gerbicz double check process verifies locally that no errors have occurred during the computation. It saves $C_1,C_2,\dots$, where $C_i=a^{2^{iB}}$ for some constant $B$ and checks that $C_{i+1}=C_i^{2^B}$ for all $i$. It then takes all required equivalences and only checks that the product of all left hand sides equal the product of all right hand sides instead of checking each individually. The Pietrzak VDF construction slightly modifies this process to make a sound, verifiable proof.
	
	The Gerbicz-Pietrzak proof scheme creates a certificate of $O(\log((\log n)/B))$ residues and takes $B$ squarings to verify. Although the Pietrzak VDF was originally proved to have unconditional soundness only when the modulus is the product of two safe primes, conditional soundness assuming the low order assumption holds for all multiplicative groups \cite{cryptoeprint:2018/712}.
	
	\subsection{Our contribution}
	
	For our purposes, the Gerbicz-Pietrzak construction is not applicable due to inhomogeneous relations between checkpoints. We first present a double check method like Gerbicz did, and we then extend it to a proof scheme with a divide-and-conquer structure, completing a practical and sound proof scheme for modular exponentiation in general.
	
	The new construction can be further generalized into the verification of an individual intervals of steps of the left-to-right modular exponentiation process, allowing for further division of the Fermat probable prime test process, to the point where it becomes feasible to distribute steps across multiple computers for world-record level probable prime tests. We have incorporated our described certificate construction to version 4 of Genefer. Our process of verification has been successfully implemented and deployed on PrimeGrid, immediately doubling the total throughput.
	
	\section{Double check process}
	
	We first outline the starting point of our proof scheme, the double check process. Taking inspiration from Gerbicz, we consider the intermediate values throughout the process of left-to-right modular exponentiation. Suppose $a$ is the base, $n$ is the exponent, and $m$ is the modulus of the left-to-right modular exponentiation process that we wish to certify. Let $L$ be the length of the binary expansion of $n$, so
	\begin{align*}n=n_02^0+n_12^1+\dots+n_{L-1}2^{L-1}\end{align*}
	is the binary expansion of $n$ where $n_i\in\{0,1\}$. Left-to-right modular exponentiation can then be described as the calculation of the sequence $u_i=a^{\lfloor\frac{n}{2^i}\rfloor}$ by the recurrence
	\begin{align*}u_i=\begin{cases}
			1 & L\le i \\ u_{i+1}^2\cdot a^{n_i} & \text{otherwise}
	\end{cases}\end{align*}
	
	The result of the modular exponentiation is $a^n\equiv u_0\pmod m$. This computation requires $L$ squarings and at most $L$ multiplications by $a$, the latter of which is cheaper than a full multiplication.
	
	For any $i$ and $j$, $u_i$ must satisfy that
	\begin{align*}
		u_i = a^{\left\lfloor\frac{n}{2^i}\right\rfloor}  = a^{\left\lfloor\left\lfloor\frac{n}{2^i}\right\rfloor/2^j\right\rfloor 2^j+(\left\lfloor\frac{n}{2^i}\right\rfloor\bmod 2^j)}  = a^{\left\lfloor\frac{n}{2^{i+j}}\right\rfloor 2^j}a^{\left\lfloor\frac{n}{2^i}\right\rfloor\bmod 2^j} = u_{i+j}^{2^j}a^{\left\lfloor\frac{n}{2^i}\right\rfloor\bmod 2^j}
	\end{align*}
	
	Saving $u_0,u_{B},u_{2B},\dots$ for some constant $B$, a double check then verifies the equivalences between $u_{iB}$ and $u_{(i+1)B}$ by gathering all equivalences together and taking a product:
	
	\begin{align*}
		\prod_{i=0} u_{iB}^{w_i} & \overset{\text{?}}{=} \prod_{i=0} u_{(i+1)B}^{w_i 2^B}a^{\left(\left\lfloor\frac{n}{2^{iB}}\right\rfloor\bmod 2^B\right) w_i}                       \\
		& = \left(\prod_{i=0} u_{(i+1)B}^{w_i}\right)^{2^B}a^{\sum\limits_{i=0} \left(\left\lfloor\frac{n}{2^{iB}}\right\rfloor\bmod 2^B\right) w_i}
	\end{align*}

	Here, we additionally weigh each equivalence to the power of some random weight $w_i$, which will be used in the actual proof scheme to ensure soundness. This method reduces the cost of the local error check from $L$ squarings to approximately $B$ squarings.
	
	\section{Interactive proof scheme}
	
	An attempt to implement the above method as a proof scheme will require a prohibitive amount of bandwidth between the prover and verifier, as $u_0,u_{B},u_{2B},\dots$ will all need to be sent for a total of $L/B$ residues, amounting to several gigabytes in record-breaking primality tests. To ensure the practicality of our method, we describe a proof construction based on divide and conquer, that reduces the size of the certificate to $\log(L/B)$ residues.
	
	In the following sections, we define $S(x,y)$ to be the coefficient $u_{x+y}^{2^y}/u_x$ caused by the additional multiplications by $a$ between $u_x$ and $u_{x+y}$, namely
	$$S(x,y) =a^{\sum\limits_{i=x}^{x+y-1}n_i2^{i-x}}=a^{\left\lfloor\frac{n}{2^x}\right\rfloor\bmod 2^y}$$
	
	\subsection{Outline of our construction}
	We first describe an informal approach to our proof process to aid implementation. Let $P(i,B)$ be the \emph{claim} that $u_{i B}=u_{(i+1)B}^{2^B} S(iB,B)$. $P(i,B)$ is effectively the assertion that the step from $u_{(i+1) B}$ to $u_{B}$ is correct. For a completely valid computation, all $P$ claims are true. For the purposes of this section, we interpret claims to be multiplicative; for example, $P(a,B)^b P(c,B)^d$ is a shorthand for
	\begin{align*}u_{a B}^b u_{c B}^d=\left(u_{(a+1)B}^b u_{(c+1) B}^d\right)^{2^B}S(a B,B)^b S(c B,B)^d\end{align*}
	
	The prover initially seeks to prove $P(0,2^x)$ for some $2^x> L$. To prove $P(a,2B)$, it is sufficient to prove $P(2a,B)$ and $P(2a+1,B)$. During the interactive proof, the prover and verifier iteratively reduce the size of a claim equivalent to $P(0,2^x)$. Suppose that the prover has, for some $y$ and weights $w_i$, a claim $\mathbf{A}=\prod_{i=0}^{c-1} P(i,2^y)^{w_i}$. The expanded form of $\mathbf{A}$ is then the following claim, where these and the following unspecified products are all from $i=0$ to $c-1$:
	\begin{align*}
		\underbrace{\left[\prod u_{i 2^y}^{w_i}\right]}_{\mathbf{A}_1}={\underbrace{\left[\prod u_{(i+1) 2^y}^{w_i}\right]}_{\mathbf{A}_2}}^{2^{(2^y)}}\underbrace{\left[\prod S(i 2^y,2^y)^{w_i}\right]}_{\text{Known.}}\tag{$\mathbf{A}$}
	\end{align*}
	
	The verifier can check this claim in $2^y$ squarings. The verifier can also interact with the prover to ask for some more information to verify this claim faster. To halve the number of squarings, we decompose $\mathbf{A}$ by the parity of $i$ in $P(i,2^y)$. Define claims $\mathbf{B}$ and $\mathbf{C}$ as these two halves:
	
	$$\mathbf{B}=\prod_{i=0}^{c-1}P(2i,2^{y-1})^{w_i}\text{ and }\mathbf{C}=\prod_{i=0}^{c-1}P(2i+1,2^{y-1})^{w_i}$$
	
	Like the Pietrzak process, the verifier randomly selects $Q$, after which the prover and verifier agree on a reduction $\mathbf{A}\iff \mathbf{B}\land\mathbf{C}\iff \mathbf{A}'=\mathbf{B}\cdot\mathbf{C}^Q$. (We prove in Section 4 that, assuming the low order assumption in $\mathbb Z_m^\times$, for a security parameter $\lambda$, when $Q$ is randomly sampled from from $\mathbb N\cup[1,2^{\lambda}]$, $\mathbf{A}$ is equivalent to $\mathbf{B}\cdot\mathbf{C}^Q$ up to probability negligible in $\lambda$.)
	
	Similarly defining $\mathbf{A}'_{1,2}$, $\mathbf{B}_{1,2}$, and $\mathbf{C}_{1,2}$ as the parts $\mathbf{A}_{1,2}$ shown above, we have:
	\begin{align*}
		\underbrace{\left[\prod u_{(2i+0) 2^{y-1}}^{w_i}\right]}_{\mathbf{B}_1=\mathbf{A}_1} & ={\underbrace{\left[\prod u_{(2i+1) 2^{y-1}}^{w_i}\right]}_{\mathbf{B}_2}}^{2^{(2^{y-1})}}\left[\prod S((2i+0) 2^{y-1},2^{y-1})^{w_i}\right]\tag{$\mathbf{B}$}              \\
		\underbrace{\left[\prod u_{(2i+1) 2^{y-1}}^{w_i}\right]}_{\mathbf{C}_1}              & ={\underbrace{\left[\prod u_{(2i+2) 2^{y-1}}^{w_i}\right]}_{\mathbf{C}_2=\mathbf{A}_2}}^{2^{(2^{y-1})}}\left[\prod S((2i+1) 2^{y-1},2^{y-1})^{w_i}\right]\tag{$\mathbf{C}$}
	\end{align*}
	
	Let $\mu=\mathbf{B}_2=\mathbf{C}_1$, which is unknown to the verifier. Then, $\mathbf{A}'_1=\mathbf{A}_1\cdot \mu^Q$ and $\mathbf{A}'_2=\mu\cdot \mathbf{A}_2^Q$. These values define the reduced claim $\mathbf{A}'$:
	$$\mathbf{A}'=\prod_{i=0}^{2c-1}P(i,2^{y-1})^{w_{\left\lfloor i/2\right\rfloor}Q^{i\bmod 2}}$$
	
	When the prover provides $\mu$ to the verifier, the number of squarings necessary to verify its original claim is halved. This process is repeated as necessary. Multiplying the sizes of the claims by $B$, e.g. from $2^x$ to $B2^x$, where $B$ is the precise value for which $u_0,u_{B},u_{2B},\dots$ were stored balances the memory consumption and the proof efficiency. In this case, the recursive interaction between the prover and the verifier ends when the current claim becomes of the form $\prod_{i=0}^{c-1} P(i,B)^{w_i}$, which the verifier can check in $B$ squarings.
	\subsection{Formal process}
	
	We now present a formal description of the interaction between the prover and the verifier. Suppose the prover and the verifier have agreed beforehand on the security parameter $\lambda$, the modulus $m$, the checkpointing rate $B$, the base of the exponentiation $a$, the exponent $n$, and an integer $x$ such that $n<2^{B2^x}$. Define the formal language
	\begin{align*}
		\mathcal{L}=\left\{(b,r,t,w_0,w_1,\dots,w_{2^{x-t}-1}):\begin{gathered}1\le b,r<m, 0\le t\le x\\ r\equiv b^{2^{B2^t}}\prod_{i=0}^{2^{x-t}-1}S(iB2^t,B2^t)^{w_i}\end{gathered}\right\}
	\end{align*}
	where the congruence is modulo $m$.
	
	The prover initially seeks to prove that $(1,r,x,1)\in\mathcal{L}$ where $r\equiv a^n\pmod m$. Consider some level of the interaction between the prover and the verifier. Suppose the prover currently claims that $(b,r,t,w_0,w_1,\dots,w_{2^{x-t}-1})\in\mathcal{L}$. Then the verifier does the following:
	
	\begin{enumerate}
		\item If any of $1\le b,r<m, 0\le t\le x$ are not satisfied, the verifier returns \texttt{reject}.
		\item If $t=0$, the verifier checks that
		\begin{align*}
			r\equiv b^{2^B}\prod_{i=0}^{2^x-1}S(iB,B)^{w_i}\equiv b^{2^B}a^{\sum\limits_{i=0}^{2^x-1}w_i\left(\left\lfloor\frac{n}{2^{iB}}\right\rfloor\bmod 2^B\right)}\pmod m
		\end{align*}
		and returns \texttt{accept} or \texttt{reject} accordingly.
		\item Otherwise, the prover computes and sends to the verifier $\mu$, where
		\begin{align*}
			\mu\equiv\prod_{i=0}^{2^{x-t}-1} u_{(2i+1) B2^{t-1}}^{w_i}=b^{2^{B2^{t-1}}}\prod_{i=0}^{2^{x-t}-1}S((2i+1)B2^{t-1},B2^{t-1})^{w_i}\pmod m
		\end{align*}
		Here, $\mu$ corresponds to the same value that the prover is asked to compute and send in Section 3.1. In other words, if the step $r\leftarrow b$ corresponds to $\mathbf{A}$ from Section 3.1, then the step $r\leftarrow \mu$ corresponds to $\mathbf{B}$ and the step $\mu\leftarrow b$ corresponds to $\mathbf{C}$.
		\item The verifier computes a challenge $Q$ randomly sampled from $\mathbb N\cup[1,2^\lambda]$, and the prover and the verifier recurse on $(b^Q\mu,\mu^Qr, t-1,w_0,Qw_0,w_1,Qw_1,\dots,w_{2^{x-t}-1},Qw_{2^{x-t}-1})$ which the prover seeks to show is in $\mathcal{L}$.
	\end{enumerate}
	
	To construct a non-interactive proof -- the certificate -- of the result $r\equiv a^n\pmod m$, it suffices to replace the verifier challenges with a hash of the current state by the Fiat-Shamir heuristic and record each $\mu$ the prover would have sent.
	
	\section{Proof of soundness}

	\begin{figure}[h]
		\begin{center}
			\includegraphics{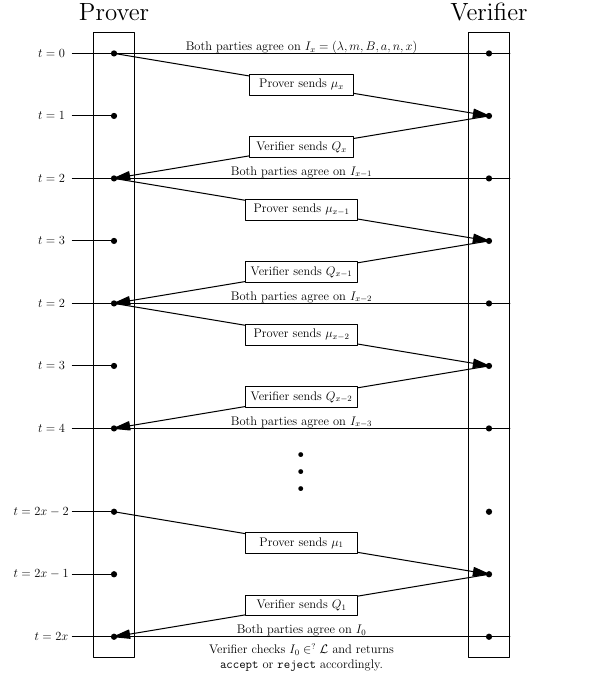}
		\end{center}
		\centering \caption{Interactive proof process}
	\end{figure}
	
	We demonstrate that for \emph{any} $(\lambda,m,B,a,n,x)$, assuming the hardness of finding an element of $\mathbb Z_m^\times$ with order less than $2^\lambda$ -- the \emph{low order assumption} -- no adversary can forge a result and proof with non-negligible probability with respect to $\lambda$.
	
	Specifically, assume the contrary; suppose there exists a randomized polynomial time adversary $\mathcal{A}$, modeled as a black box, defined as
	\begin{align*}\mathcal{A}(\lambda,m,B,a,n,x;Q_x,Q_{x-1},\dots,Q_1)\to I_x,(\mu_x, I_{x-1}),(\mu_{x-1},I_{x-2}),\dots,(\mu_1,I_0)\end{align*}
	
	When given $(\lambda,m,B,a,n,x)$ and randomly sampled $Q_x,Q_{x-1},\dots,Q_1\leftarrow \mathbb N\cup[1,2^\lambda]$, $\mathcal{A}$ attempts to generate an input $I_x=(1,r,x,1)$ and an corresponding interaction \begin{align*}(\mu_x, I_{x-1}),(\mu_{x-1},I_{x-2}),\dots,(\mu_1,I_0)\end{align*} (see figure 1) and \emph{succeeds} with probability non-negligible in $\lambda$.
	
	We say that $\mathcal{A}$ \emph{succeeds} if and only if
	
	\begin{enumerate}
		\item $I_x\not\in\mathcal{L}$ and $I_0\in\mathcal{L}$, i.e. $\mathcal{A}$ deceives a verifier with challenges $Q_x,Q_{x-1},\dots,Q_1$.
		\item For all $y$ and all $Q'_y,Q'_{y-1},\dots,Q_1\leftarrow \mathbb N\cup[1,2^\lambda]$, we have that when $\mathcal{A}$ is run with the same random tape (i.e. makes the same random decisions) it must hold that $I_y=I'_y$ and $\mu_y=\mu'_y$, where
		{\small\begin{align*}
				\mathcal{A}(\lambda,m,B,a,n,x;Q_x,\dots,Q_{y+1},Q_y,\dots,Q_1)   & \to I_x,(\mu_x, I_{x-1}),(\mu_{x-1},I_{x-2}),\dots,(\mu_1,I_0)        \\
				\mathcal{A}(\lambda,m,B,a,n,x;Q_x,\dots,Q_{y+1},Q'_y,\dots,Q'_1) & \to I'_x,(\mu'_x, I'_{x-1}),(\mu'_{x-1},I'_{x-2}),\dots,(\mu'_1,I'_0)
		\end{align*}}
	\end{enumerate}
	
	The latter condition is necessary to ensure that $\mathcal{A}$ does not ``look ahead'' and base decisions (of $I$ and $\mu$) based on future challenges. We now prove that the success of $\mathcal{A}$ contradicts the low-order assumption, and thus the probability of forging a proof is negligible with respect to $\lambda$.
	
	\newtheorem{thm}{Theorem}
	
	\begin{thm}
		For some $(\lambda,m,B,a,n,x)$, if $\mathcal{A}$ succeeds with probability non-negligible with respect to $\lambda$, there exists an adversary that can use $\mathcal{A}$ twice to obtain an element of $\mathbb Z_m^\times$ with order less than $2^{\lambda}$ with non-negligible probability, breaking the low-order assumption.
	\end{thm}
	
	To this end we will prove two claims:
	
	\begin{enumerate}
		\item If, using $\mathcal{A}$, an adversary finds a state $I_y\not\in\mathcal{L}$, a prover message $\mu_y$, and two separate challenges $Q_y,Q'_y$ such that the resulting states $I_{y-1}$ and $I'_{y-1}$ are both in $\mathcal{L}$, then it can quickly recover some element $E\not\equiv 1\pmod m$ and some exponent $0<r<2^{\lambda}$ such that $E^r\equiv 1\pmod m$.
		\item If $\mathcal{A}$ succeeds with probability $P$, an adversary can use $\mathcal{A}$ twice and succeed in finding the aforementioned $(I_y,\mu_y,Q_y,Q'_y)$ with probability at least $P(P/x-2^{-\lambda})$.
	\end{enumerate}
	
	\begin{proof}[Proof of first claim]
		When the adversary indeed succeeds in finding $(I_y,\mu_y,Q_y,Q'_y)$, we have some $I_y$ such that $I_y\not\in\mathcal{L}$ and two $I_{y-1}, I'_{y-1}$, caused by $Q_y$ and $Q'_y$ respectively, such that $I_{y-1},I'_{y-1}\in\mathcal{L}$.
		
		For a state $I=(b,r,t,w_0,w_1,\dots,w_{2^{x-t}-1})$, define $R(I)=r$, $B(I)=b$, and $C(I)$ as
		\begin{align*}C(I)=\sum_{i=0}^{2^{x-t}-1}w_i\left(\left\lfloor\frac{n}{2^{iB2^t}}\right\rfloor\bmod 2^{B2^t}\right)
		\end{align*}
	
		If we define $c_1$ and $c_2$ as
		{\begin{align*}
				c_1 =\sum\limits_{i=0}^{2^{x-y}-1}w_i\left(\left\lfloor\frac{n}{2^{(2i+0)B2^{y-1}}}\right\rfloor\bmod 2^{B2^{y-1}}\right) \text{ and }
				c_2 =\sum\limits_{i=0}^{2^{x-y}-1}w_i\left(\left\lfloor\frac{n}{2^{(2i+1)B2^{y-1}}}\right\rfloor\bmod 2^{B2^{y-1}}\right)
		\end{align*}}
		so $C(I_y)=c_1+2^{B2^{y-1}}c_2$, we have
		\begin{align*}
			 B(I_y)^{2^{B2^y}}a^{C(I_y)}=\left(B(I_y)^{2^{B2^{y-1}}}\right)^{2^{B2^{y-1}}} a^{c_1+2^{B2^{y-1}}c_2}= \left(B(I_y)^{2^{B2^{y-1}}}a^{c_2}\right)^{2^{B2^{y-1}}} a^{c_1}
		\end{align*}

		By $I_y\not\in\mathcal{L}$ we have, where this and the following equivalences are all modulo $m$:
		\begin{align*}
			R(I_y)\not\equiv \left(B(I_y)^{2^{B2^{y-1}}}a^{c_2}\right)^{2^{B2^{y-1}}} a^{c_1}
		\end{align*}
		
		Thus, at least one of the following does not hold:
		\begin{align*}
			R(I_y) & \equiv \mu_y^{2^{B2^{y-1}}}a^{c_1}\tag{1} \\
			\mu_y  & \equiv B(I_y)^{2^{B2^{y-1}}}a^{c_2}\tag{2}
		\end{align*}
		
		By $I_{y-1},I'_{y-1}\in\mathcal{L}$, for $I_{y-1}$ have $R(I_{y-1})\equiv B(I_{y-1})^{2^{B2^{y-1}}}a^{C(I_{y-1})}$. At the same time, our recursion $I_y\to I_{y-1}$ is defined with $B(I_{y-1})=B(I_y)^Q\mu_y$, $R(I_{y-1})=\mu_y^QR(I_y)$, and $C(I_{y-1}) = c_1+Qc_2$; expanding for $I_{y-1}$ and $I'_{y-1}$ gives
		\begin{align*}
			\mu_y^QR(I_y)    & \equiv \left(B(I_y)^Q\mu_y\right)^{2^{B2^{y-1}}}a^{c_1+Qc_2}     \\
			\mu_y^{Q'} R(I_y) & \equiv \left(B(I_y)^{Q'}\mu_y\right)^{2^{B2^{y-1}}}a^{c_1+Q'c_2}
		\end{align*}
		
		Rearranging,
		\begin{align*}
			R(I_y)/\left(\mu_y^{2^{B2^{y-1}}}a^{c_1}\right) \equiv \left(B(I_y)^{2^{B2^{y-1}}}a^{c_2}/\mu_y\right)^Q \equiv \left(B(I_y)^{2^{B2^{y-1}}}a^{c_2}/\mu_y\right)^{Q'} \tag{3}
		\end{align*}

		If $(1)$ is false, then $R(I_y)/\left(\mu_y^{2^{B2^{y-1}}}a^{c_1}\right)\not\equiv 1$ but $\left(B(I_y)^{2^{B2^{y-1}}}a^{c_2}/\mu_y\right)^Q$, which is equivalent to $1$ if $(2)$ is true, is equivalent to $R(I_y)/\left(\mu_y^{2^{B2^{y-1}}}a^{c_1}\right)$ by $(3)$ and thus not equivalent to $1$; therefore $(2)$ is false.

		It follows that in any case $(2)$ must be false. $(3)$ then gives that $E^{|Q-Q'|}\equiv 1$ where
		\begin{align*}
			E=B(I_y)^{2^{B2^{y-1}}}a^{c_2}/\mu_y\not\equiv 1
		\end{align*}
		and $1\le|Q-Q'|< 2^{\lambda}$.
		
		This demonstrates that in the event of the adversary indeed finding $(I_y,\mu_y,Q_y,Q'_y)$, an element $E\not\equiv 1$ is also found with order less than $2^\lambda$.
	\end{proof}
	
	It remains to analyze the probability of the adversary finding $(I_y,\mu_y,Q_y,Q'_y)$, the desired setting, with only two uses of $\mathcal{A}$.
	
	\begin{proof}[Proof of second claim]
		We leverage the forking argument used in ``A Survey of Two Verifiable Delay Functions'' \cite{cryptoeprint:2018/712}. Specifically, we reinterpret $\mathcal{A}$ as part of a new process $\mathcal{A}'$ that is more amenable to the generalized forking lemma introduced by Bellare and Neven \cite{10.1145/1180405.1180453}.
		
		Let us abstract $\mathcal{A}$ as a probabilistic Turing machine with random tape $R$. Define \begin{align*}\mathcal{A'}(\lambda,m,B,a,n,x;Q_x,Q_{x-1},\dots,Q_1;R)\end{align*} to represent an execution of $\mathcal{A}$ with the given parameters and random tape $R$; $\mathcal{A'}$ returns $(\epsilon,\epsilon,\epsilon,\epsilon)$ if $\mathcal{A}$ fails, and otherwise outputs $(y,I_y,\mu_y,I_{y-1})$ where $y=\arg\min(y:I_y\not\in\mathcal{L})$.
		
		For a given $(\lambda,m,B,a,n,x)$, the adversary then proceeds as follows:
		
		\begin{enumerate}
			\item It randomly samples $Q_x,Q_{x-1},\dots,Q_1\leftarrow \mathbb N\cup[1,2^\lambda]$ and generates a random tape $R$.
			\item It executes $y,I_y,\mu_y,I_{y-1}\leftarrow \mathcal{A}'(\lambda,m,B,a,n,x;Q_x,Q_{x-1},\dots,Q_1;R)$. If $y=\epsilon$ the adversary fails.
			\item It randomly samples $Q'_y,Q'_{y-1},\dots,Q'_1\leftarrow \mathbb N\cup[1,2^\lambda]$.
			\item It executes
			\begin{align*}y',I'_y,\mu'_y,I'_{y-1}\leftarrow \mathcal{A}'(\lambda,m,B,a,n,x;Q_x,Q_{x-1},\dots,Q_{y+1},Q'_y,Q_{y-1},\dots,Q'_1;R)\end{align*}
			and if $y'=\epsilon$ or $y'\neq y$ the adversary fails.
			\item By the second condition for success of $\mathcal{A}$, it now must hold that $I'_y=I_y$ and $\mu'_y=\mu_y$, as neither the random tape nor $Q_{1\dots x}$ have changed.
			\item If $Q_y\neq Q'_y$, it succeeds and generates $(y,I_y,\mu_y,Q_y,Q'_y,I_{y-1},I'_{y-1})$; otherwise, it fails.
		\end{enumerate}
		
		By the generalized forking lemma, if $\mathcal{A}$ succeeds with probability $P$, then the adversary succeeds with probability of at least $P(P/x-1/2^{\lambda})$. It follows that if $\mathcal{A}$ succeeds with non-negligible probability, then the adversary also succeeds with non-negligible probability, as $x\in O(\log\log m)\in O(\operatorname{poly}\lambda)$.
	\end{proof}
	
	We have shown that if $(I_y,\mu_y,Q_y,Q'_y)$ is found with non-negligible probability with respect to $\lambda$ the low order assumption is broken, and if the adversary succeeds with non-negligible probability then $(I_y,\mu_y,Q_y,Q'_y)$ is found with non-negligible probability. 
	
	Therefore, combined, these two lemmas complete the desired proof of conditional soundness for our modular exponentiation proof scheme. If the low order assumption holds in $\mathbb Z_m^\times$, then for any initial configuration $(\lambda,m,B,a,n,x)$, the probability of a forged proof convincing the verifier for a state $I\not\in\mathcal{L}$ is negligible with respect to $\lambda$.
	
	\section{Final remarks}
	
	\subsection{Implementation and time-space tradeoff}
	
	A direct implementation of the formal definition of our proof scheme defined in Section 3.2 is impractical as it requires the calculation of $b^{2^{B2^{t-1}}}$ for $t=x\to 1$, incurring an additional $L$ squarings; exactly what we want to avoid. Using the abstraction described in Section 3.1, we see that it is equivalent to instead store $b^{2^{Bi}}$ for $i=0\dots 2^x-1$; however, for smaller $B$, the computation becomes bottlenecked by disk I/O. In distributed computing, this entails balancing $B$ not only between the additional prover cost and the verifier, but also the disk space the prover has available.
	
	On the other hand, we can reverse the time-space tradeoff, by instead choosing a larger $B$ for the prover. This does not necessarily mean verification becomes more expensive; the final bottleneck for the verifier is always the verification of $r\equiv b^{2^B}a^c$, which takes polynomial (in $L$) squarings, as opposed to the interaction, which takes a polylogarithmic number of squarings. This is exactly the initial form, i.e. the prover needs that $(b,r,x',1)\in\mathcal{L}'$ where $\mathcal{L}'$ is defined as in Section 3.2 with the same $a$, the same $m$, some different $B'$ and $x'$ such that $B'2^{x'}=B$, and a new $n=c$. In other words, the prover and verifier can recurse again on $(b,r,x',1)$ and $\mathcal{L}'$, reducing the number of squarings for the verifier to $B'$, at the cost of $B$ extra squarings from the prover. Iterating this meta-process delegates all of the work to the prover, creating a proof that can be verified in polylogarithmic in $\log m$ time.
	
	\subsection{Implications of conditional soundness}
	
	Due to Shor's algorithm, the low-order assumption does not hold in the quantum computing model \cite{365700}, as there exists a quantum adversary that computes the discrete logarithm in $\tilde O(\left(\log m\right)^2)$. This does not yet pose a practical threat to distributed computing purposes: for probabilistic primality tests, not only is $\log m\gg 10^5$ several orders of magnitude larger than the intended scale of Shor's algorithm for decrypting public-key cryptosystems, but because $n<m$ for the purpose of primality tests, classical algorithms also run in $\tilde O(\left(\log m\right)^2)$ and likely much faster. This \emph{de facto} safety may change if our results are used to optimize other applications, such as distributed exponentiation.

	\subsection{Implementation at PrimeGrid}
	
	Our proof scheme and its corresponding verification process has been successfully implemented at PrimeGrid, doubling the speed of PrimeGrid's search that spans hundreds of thousands of computers worldwide. Among others, the search for the world's largest prime on PrimeGrid now uses our proof scheme.
	
	The first prime found using our work is $117687318^{131072}+1$, with $1057847$ digits, discovered by Tom Greer.
	
	\section{References}
	
	\printbibliography[heading=none] 

	\section{Acknowledgements}
	
	The author thanks his advisor, Yves Gallot, not only for his invaluable advice for this paper, but also for his continued development of Genefer and its contribution to PrimeGrid, without either of which the author would never have realized the application of number theory to distributed computing, as well as the PrimeGrid administrators, for their rapid integration and diligent testing of our implementation into the PrimeGrid servers.
	
	The author thanks the Open Science Grid (OSG) for providing CPU and GPU computational resources for the testing of our proof scheme at PrimeGrid using the new version of Genefer. This research was done using services provided by the OSG Consortium \cite{osg07}\cite{osg09}, which is supported by the National Science Foundation awards \#2030508 and \#1836650.

	\vfill
	
	\begin{center}
		\emph{this result is dedicated to Ruvim Breydo, who taught me how to do real math}
	\end{center}
	
\end{document}